\documentclass[aps,pra,twocolumn,showpacs,amsmath,amssymb,floatfix,footinbib,superscriptaddress]{revtex4-1}
\usepackage{amsmath,natbib,graphicx,subfigure,amssymb,graphics,amsmath,mathrsfs,CJK,color}
\usepackage{multirow,fancyhdr,color,bm,tabularx,psfrag,dcolumn}
\usepackage[colorlinks=true,linkcolor=blue,urlcolor=blue,citecolor=blue]{hyperref}
\usepackage{tikz,xcolor,hyperref}
\definecolor{lime}{HTML}{A6CE39}
\DeclareRobustCommand{\orcidicon}{%
    \begin{tikzpicture}
    \draw[lime, fill=lime] (0,0)
    circle [radius=0.16]
    node[white] {{\fontfamily{qag}\selectfont \tiny ID}};\draw[white, fill=white] (-0.0625,0.095)
    circle [radius=0.007];
    \end{tikzpicture}
    \hspace{-2mm}}
\foreach \x in {A, ..., Z}
{\expandafter\xdef\csname orcid\x\endcsname{\noexpand\href{https://orcid.org/\csname orcidauthor\x\endcsname}{\noexpand\orcidicon}}}

\usepackage[left]{lineno}
\begin{document}

\title{ 2-D isotropic negative refractive index in a N-type four-level atomic system }
\thanks{ Supported by National Natural Science Foundation of China (NSFC) (Grant Nos.61205205 and 6156508508)
and the Foundation for Personnel training projects of Yunnan Province(grant No.KKSY201207068) of
China.\\ $^{\dag}$Corresponding author: zscnum1@126.com, zscgroup@kmust.edu.cn }

\author{Shun-Cai Zhao\orcidA{}}
\email[Corresponding author: ]{zhaosc@kmust.edu.cn.}
\affiliation{Faculty of science, Kunming University of Science and Technology, Kunming, 650093, PR China}
\affiliation{Center for Quantum Materials and Computational Condensed Matter Physics, Faculty of Science, Kunming University of Science and Technology, Kunming, 650500, PR China}

\author{Qi-Xuan Wu}
\affiliation{College English department, Kunming University of Science and Technology, Kunming, 650500, PR China}

\author{Kun Ma}
\affiliation{Faculty of science, Kunming University of Science and Technology, Kunming, 650093, PR China}
\affiliation{Center for Quantum Materials and Computational Condensed Matter Physics, Faculty of Science, Kunming University of Science and Technology, Kunming, 650500, PR China}
 
\begin{abstract}
2-D(Two-dimensional) isotropic negative refractive index (NRI) is explicitly
realized via the orthogonal signal and coupling standing-wave fields coupling
the N-type four-level atomic system. Under some key parameters of the dense
vapor media, the atomic system exhibits isotropic NRI with simultaneous
negative permittivity and permeability (i.e. Left-handedness) in the 2-D x-y plane.
Compared with other 2-D NRI schemes, the coherent atomic vapor media in our scheme
may be an ideal 2-D isotropic NRI candidate and has some potential
advantages, significance or applications in the further investigation.
\end{abstract}

\maketitle

\section{Introduction}
J.B. Pendry, et al. asserted that a medium with negative effective
permeability can be achieved by the split-ring resonator (SRR) in free
space\cite{1}, which has been studied to increase the permeability of artificial
dielectrics in the 1952s\cite{2}. Smith et al. combined the SRR with thin wires to
design the NRI medium in 2000\cite{3}. And a year later, this SRR/wire
composite medium was implemented and experimentally verified negative refraction for the first
time\cite{4}. The implementation has given rise to the NRI media because of its exotic properties.
In particular, the possible sub-wavelength resolving properties\cite{1} have stirred much excitement
and renewed interest in electromagnetic phenomena associated with NRI medium, first investigated by
Veselago in the 1960s\cite{5}.

With the development of investigating the NRI media, the requirement of 2-D NRI media is emerging.
And several schemes\cite{6,7,8,9,10,11} for 2-D NRI media have been proposed recently.
However, most of the proposed 2-D NRI media are based on photonic crystal nanostructure or transmission
line structure, and most of them are anisotropic, i.e., their exotic properties are polarization dependent,
which is undesirable in certain potential applications, e.g., in ``perfect lens''\cite{1}. As an
alternative, the arbitrary linear incident polarization via the printable NRI media have been proposed
by C. Imhof et al..\cite{12}.

Here, we first explore the possibility of implementing 2-D NRI media by using
the quantum coherence effect in a four-level atomic system, which is driven by
the orthogonal signal and coupling standing-wave fields. Thus, we can explicitly
investigate the left-handedness and NRI of the resonant atomic system in the x-y plane.
A few specific NRI values exhibit circular and homogeneous distributions in the x-y plane,
and the atomic system possesses negative permittivity and permeability simultaneously.

\section{Model and equation}
\begin{figure}[htp]
\center
\includegraphics[totalheight=1.6 in]{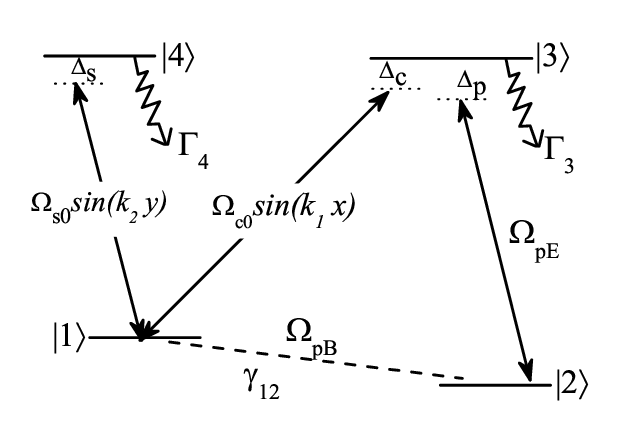}
\caption{Schematic diagram of a four-level N-type atomic system. The
electric-dipole transitions $|1\rangle$-$|4\rangle$
and$|1\rangle$-$|3\rangle$ are driven by two orthogonal standing-wave fields
$\Omega_{c}(x)$ and $\Omega_{s}(y)$, respectively. The level
pairs $|3\rangle$-$|2\rangle$ and$|1\rangle$-$|2\rangle$ are coupled
to the electric and magnetic components of the weak probe light,
respectively. $\gamma_{3,4}$ and $\gamma_{12}$ denote the
spontaneous decay rates and the dephasing rate, respectively.
\label{1}}
\end{figure}
Consider a N-type four-level atomic system with two lower states
$|1\rangle$ and $|2\rangle$, two excited states $|3\rangle$ and
$|4\rangle$, which interacts with three optical fields, i.e., the
weak probe light $\Omega_{p}$ and two strong coherent coupling standing-wave
fields $\Omega_{c}(x)$, $\Omega_{s}(y)$, respectively [see Fig.1].
Levels $|1\rangle$ and $|2\rangle$ have the same parity and level
$|2\rangle$, $|3\rangle$ with an opposite parity. Thus the electric
and magnetic components of the weak probe light field couple the electric-dipole transition $|3\rangle$-$|2\rangle$ and
the magnetic-dipole transition $|1\rangle$-$|2\rangle$,
respectively. Such the atomic system may be realized in
$^{171}Yb$(I=1/2). The states $|1\rangle$, $|2\rangle$, $|3\rangle$
and $|4\rangle$ correspond to $|^{1}S_{0}$, F=1/2,$
m_{F}$=-1/2$\rangle$, $|^{1}S_{0}$,F=1/2, $m_{F}$=1/2$\rangle$,
$|^{1}P_{1}$, F=1/2, $m_{F}$=-1/2$\rangle$ and $|^{3}P_{1}$, F=1/2,
$m_{F}$=1/2$\rangle$, respectively.

The Rabi frequencies corresponding to the two
orthogonal standing-wave fields are
$\Omega_{c}(x)$=$d_{13}E_{c}/\hbar$=$\Omega_{c0}sin(k_{1}x)$,
$\Omega_{s}(y)$=$d_{14}E_{c}/\hbar$=$\Omega_{s0}sin(k_{2}y)$, respectively.
$E_{c}$, $E_{s}$ are the electric field strength envelopes of the
two orthogonal standing-wave fields. $k_{i}$=$2\pi/\lambda_{i}$, (i
= 1, 2) is the wave vector with wavelengths $\lambda_{i}$, (i= 1, 2)
of the corresponding standing wave fields. And the two orthogonal
signal and coupling standing-wave fields drive the
transitions $|1\rangle$-$|4\rangle$ and$|1\rangle$-$|3\rangle$,
respectively. The frequency detunings of these three optical fields
are $\Delta_{p}$, $\Delta_{c}$ and $\Delta_{s}$, respectively. At a
low temperature, the atoms are assumed to be nearly stationary
and hence any Doppler shift is neglected. Under the rotating-wave approximations the equations of motion for
the probability amplitudes of the atomic system in the x-y plane are given as follows,
\vskip -0.7cm
\begin{eqnarray}
&\frac{\partial A_{1}(x,y)}{\partial t}=&\frac{i}{2}\left(\Omega^{*}_{pE}A_{3}(x,y)+\Omega^{*}_{pB}A_{2}(x,y)\right),\nonumber\\
&\frac{\partial A_{2}(x,y)}{\partial t}=&-i(\Delta_{p}-\Delta_{c})A_{2}(x,y)+\frac{i}{2}(\Omega^{*}_{c}(x)A_{3}(x,y) \nonumber \\
                                         && +\Omega^{*}_{s}(y)A_{4}(x,y))+\frac{i}{2}\Omega^{*}_{pB}A_{1}(x,y)\nonumber \\
                                         &&-\frac{\gamma_{12}}{2}A_{2}(x,y), \label{Eq1}\\
&\frac{\partial A_{3}(x,y)}{\partial t}=&-i\Delta_{p}A_{3}(x,y)+\frac{i}{2}(\Omega_{c}(x)A_{2}(x,y)\nonumber\\
                                        &&+\Omega_{p}A_{1}(x,y))-\frac{\Gamma_{3}}{2}A_{3}(x,y),\nonumber\\
&\frac{\partial A_{4}(x,y)}{\partial t}=&-i\left(\Delta_{p}+\Delta_{s}-\Delta_{c}\right)A_{4}(x,y)+\frac{i}{2}\Omega_{s}(y) \nonumber\\
                                         && A_{2}(x,y)-\frac{\Gamma_{4}}{2}A_{4}(x,y) \nonumber
\end{eqnarray}
\vskip -0.6cm
where the electric and magnetic Rabi frequencies of the probe light
are defined by $\Omega_{pE}=d_{23}E_{p}/\hbar$,
$\Omega_{pB}=\mu_{12}B_{p}/\hbar$, respectively. Here, $E_{p}$ and
$B_{p}$ denote the electric and magnetic field strength
envelopes of the probe light. The frequency detunings are defined by
$\Delta_{p}$=$\omega_{32}-\omega_{p}$,
$\Delta_{c}$=$\omega_{31}-\omega_{c}$, and
$\Delta_{s}$=$\omega_{41}-\omega_{s}$. Where $\omega_{p}$,
$\omega_{c}$ and $\omega_{s}$ are the mode frequencies of the probe
and two orthogonal standing-wave fields. $\Gamma_{3,4}$ and
$\gamma_{12}$ in Eq.(1) denote the spontaneous decay rates and the
dephasing rate, respectively.

When the intensity of the probe light is sufficiently
weak and the temperature is very low, all the atoms remain in the ground state and the atomic
population in level $|3\rangle$ is close to unity\cite{13} without any Doppler shift. Under
these conditions, the steady solutions of  Eq. (1) take the
following forms
\vskip -0.7cm
\begin{eqnarray*}
&A_{2}(x,y)=&\frac{1}{\xi(x,y)}[\frac{\gamma_{4}}{2}+i(\Delta_{s}+\Delta_{p}-\Delta_{c})] [\frac{\Omega^{*}_{c}(x)\Omega_{pE}}{4} \nonumber\\
                  &&-\frac{i}{2}\cdot(\frac{\gamma_{3}}{2}+i\Delta_{p})\Omega_{pB}],\\
\end{eqnarray*}
\begin{eqnarray}
&A_{3}(x,y)=&-\frac{i}{2\xi(x,y)}[\frac{\gamma_{4}}{2}+i(\Delta_{s}+\Delta_{p} -\Delta_{c})]\{[\frac{\gamma_{12}}{2}     \nonumber\\
                  &&+i(\Delta_{p}-\Delta_{c})]\Omega_{pE}+\frac{i}{2}\Omega_{c}(x)\Omega_{pB}\}-\frac{i}{8\xi(x,y)}      \nonumber\\
                  &&\Omega_{s}(y)\Omega^{*}_{s}(y)\Omega_{pE},\label{Eq2} \\
&A_{4}(x,y)=&-\frac{i}{2\xi(x,y)}[-\frac{1}{4}\Omega_{pE}\Omega^{*}_{c}(x)+\frac{i}{2}(\frac{\gamma_{3}}{2}+i\Delta_{p})\Omega_{pB}]\Omega_{s}(y),\nonumber
\end{eqnarray}
with the parameter
\begin{eqnarray*}
&\xi(x,y)=&-\{\frac{\Omega_{c}(x)\Omega^{*}_{c}(x)}{4}+[\frac{\gamma_{12}}{2}+i(\Delta_{p}-\Delta_{c})](\frac{\gamma_{3}}{2}+i\Delta_{p})\}\\
&&[\frac{\gamma_{4}}{2}+i(\Delta_{s}+\Delta_{p}-\Delta_{c})]-\frac{\Omega_{s}(y)\Omega^{*}_{s}(y)}{4}(\frac{\gamma_{3}}{2}+i\Delta_{p}),\nonumber\\
\end{eqnarray*}\vskip -0.7cm
In the following, we consider the electric and magnetic responses of
the N-type four-level atomic system interacting with two orthogonal
stand-wave light fields.

Considering the distinction between the applied fields and the
microscopic local fields acting upon the atoms when discussing the
detailed properties of the atomic transitions between the levels
relating to the electric and magnetic susceptibilities, there is
little difference between the macroscopic fields and the local
fields\cite{14,17} in the dilute vapor case. However, the polarization
of neighboring atoms gives rise to an internal field at any given
atom in addition to the average macroscopic field. In order to
achieve the negative permittivity and permeability one should
consider the local field effect in dense media with closely packed
atom.

We first obtain the atomic electric and magnetic
polarizabilities, and then consider the local field correction to
the electric and magnetic susceptibilities of the coherent vapor
medium. Using the definitions of atomic microscopic electric and
magnetic polarizabilities\cite{14,15}, we can arrive at the following
parameters
\begin{eqnarray}
&\gamma_{e}(x,y)=&\frac{id^{2}_{23}}{\hbar\varepsilon_{0}}\frac{\frac{\Omega_{s}(y)\Omega^{*}_{s}(y)}{4}+[\frac{\gamma_{12}}{2}+i(\Delta_{p}-\Delta_{c})+\frac{i}{2}\frac{\mu_{12}}{cd_{23}}\Omega_{c}(x)]}{\{\frac{\Omega_{c}(x)\Omega^{*}_{c}(x)}{4}+[\frac{\gamma_{12}}{2}+i(\Delta_{p}-\Delta_{c})](\frac{\gamma_{3}}{2}+i\Delta_{p})\}}\nonumber\\
            &&\frac{[\frac{\gamma_{4}}{2}+i(\Delta_{s}+\Delta_{p}-\Delta_{c})]}{[\frac{\gamma_{4}}{2}+i(\Delta_{s}+\Delta_{p}-\Delta_{c})]+\frac{\Omega_{s}(y)\Omega^{*}_{s}(y)}{4}(\frac{\gamma_{3}}{2}+i\Delta_{p})},
\end{eqnarray}
\begin{eqnarray}
&\gamma_{m}(x,y)=&\frac{-2\mu_{0}\mu^{2}_{12}}{\hbar}\frac{[\frac{\gamma_{4}}{2}+i(\Delta_{s}+\Delta_{p}-\Delta_{c})][\frac{\Omega^{*}_{c}(x)}{4}\frac{cd_{23}}{\mu_{12}}}
{{\{\frac{\Omega_{c(x)}\Omega^{*}_{c}(x)}{4}+[\frac{\gamma_{12}}{2}+i(\Delta_{p}-\Delta_{c})]}}                        \nonumber\\
&&\frac{-\frac{i}{2}(\frac{\gamma_{3}}{2}+i\Delta_{p})]}{(\frac{\gamma_{3}}{2}+i\Delta_{p})\}[\frac{\gamma_{4}}{2}+i(\Delta_{s}+\Delta_{p}-\Delta_{c})]}\nonumber\\
&&\frac{}{+\frac{\Omega_{s}(y)\Omega^{*}_{s}(y)}{4}(\frac{\gamma_{3}}{2}+i\Delta_{p})}
\end{eqnarray}
In the above calculations,
$\gamma_{e}(x,y)$ = $\frac{2d^{2}_{23}A_{3}(x,y)}{\hbar\varepsilon_{0}\Omega_{pE}}$,
$\gamma_{m}(x,y)$ = $\frac{2\mu_{0}\mu_{12}A_{2}(x,y)}{B_{p}}$ and the relation
between the microscopic local electric and magnetic fields
$E_{p}/B_{p}$ =$ c $ has been inserted.

We have obtained the microscopic physical quantities $\gamma_{e,m}(x,y)$.
However, what we are interested in is the macroscopic physical
quantities such as the electric permittivity and magnetic
permeability. According to both the electric and magnetic
Clausius-Mossotti relations\cite{14,15}, we can arrive at the
relative electric permittivity $\varepsilon_{r}(x,y)$ and magnetic
permeability $\mu_{r}(x,y)$ as follows
\begin{eqnarray}
\varepsilon_{r}(x,y)=\frac{1+\frac{2}{3}N\gamma_{e}(x,y)}{1-\frac{1}{3}N\gamma_{e}(x,y)}, \nonumber\\
\mu_{r}(x,y)=\frac{1+\frac{2}{3}N\gamma_{m}(x,y)}{1-\frac{1}{3}N\gamma_{m}(x,y)}\nonumber
\end{eqnarray}
where N denotes the atomic concentration.

\section{Results and discussions}

Here we will demonstrate the 2-D isotropic NRI can truly
arise in the four-level coherent atomic vapor media interacting
with two orthogonal standing-wave fields in the x-y plane.
And the strong electric and magnetic responses can bring
about simultaneously negative permittivity and permeability.
For the sake of more closing to reality, several key
parameters are adopted from the form investigation. The
parameters for the atomic electric and magnetic polarizability
are chosen as: the electric and magnetic transition dipole moments
$d_{23}$=$3\times10^{-29}C \cdot m $ and $\mu_{12}=1.3\times10^{-22}
C\cdot m^{2}s^{-1} $ \cite{16}, respectively. The density of atom N
was chosen to be $2\times10^{23}$ $ m^{-3}$ \cite{17}. And the other
parameters are scaled by $\gamma=10^{8}s^{-1}$: $\Gamma_{3}=0.3\gamma$,
$\Gamma_{4}=0.1\gamma$. Because the dephasing rate is in general two or
three orders of magnitude less than the spontaneous decay rates in a low-temperature vapor
\cite{18}, the dephasing rate $\gamma_{12}$ is small and set
to be $10^{-3}\gamma$.

\begin{figure}[htp]
\center
\includegraphics[totalheight=1.0 in]{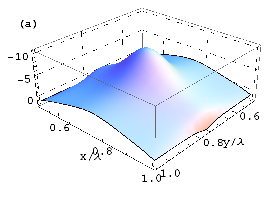}\includegraphics[totalheight=1.0 in]{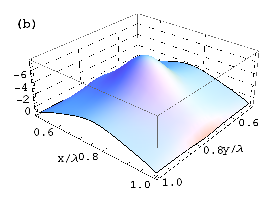}
\hspace{0in}%
\includegraphics[totalheight=1.0 in]{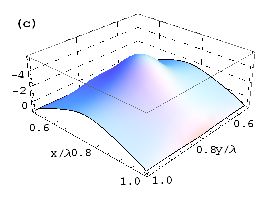}\includegraphics[totalheight=1.0 in]{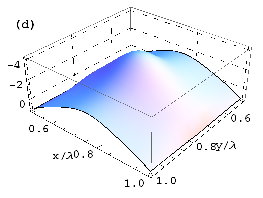}
\caption{(Color online) Plots for the 2-D $Re\{\varepsilon_{r}\}$ as a function of (x,y)
dependent the probe detuning $\Delta_{p}$. (a)
$\Delta_{p}$=$4.7\gamma$, (b) $\Delta_{p}$=5.0$\gamma$, (c)
$\Delta_{p}$=5.3$\gamma$, (d) $\Delta_{p}$=5.7$\gamma$. The other
parameters are $\Omega_{c0}$=10.2$\gamma$,
$\Omega_{s0}$=9.5$\gamma$, $\Delta_{c}$=$-\Delta_{s}$=$-0.15\gamma$,
$\gamma_{3}$=$0.3\gamma$ and $\gamma_{4}$=$0.1\gamma$,
$\gamma$=$10^{8}$$s^{-1}$.\label{2}}
\end{figure}
\begin{figure}[htp]
\center
\includegraphics[totalheight=1.0 in]{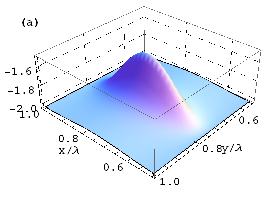}\includegraphics[totalheight=1.0 in]{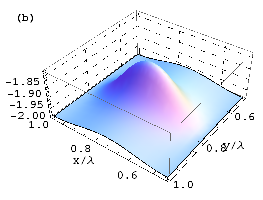}
\hspace{0in}%
\includegraphics[totalheight=1.0 in]{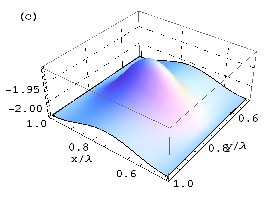}\includegraphics[totalheight=1.0 in]{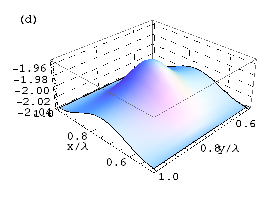}
\caption{(Color online) Plots for the 2-D $Re\{\mu_{r}\}$ as a function of (x,y)
dependent the probe detuning $\Delta_{p}$. All the parameters in (a) to (d)
are the same as in Figs. 2(a) to 2(d), respectively.\label{3}}
\end{figure}
Then the three-dimensional plots depict the real part of relative permittivity $Re\{\varepsilon_{r}\}$,
permeability $Re\{\mu_{r}\}$ and refraction index $Re\{n\}$ of the atomic system dependent
the probe detuning $\Delta_{P}$ as a function of (x, y), which are shown in Figs.2, 3 and 4, respectively.
As shown in Fig.2, the real part of relative
permittivity $Re\{\varepsilon_{r}\}$ has the negative value in the
x-y plane and the spike-like peak at the coordinate
$\{0.75\lambda,0.75\lambda\}$ in Fig.2(a), Fig.2(b), Fig.2(c), Fig.2(d),
respectively. We also noted $Re\{\varepsilon_{r}\}$ remains the negative values
with the decreasing peaks when the values of $\Delta_{p}$
are tuned to be $5.0\gamma$, $5.3\gamma$ and $5.7\gamma$, in Fig.2(b), Fig.2(c), Fig.2(d), respectively.

\begin{figure}[htp]
\center
\includegraphics[totalheight=1.0 in]{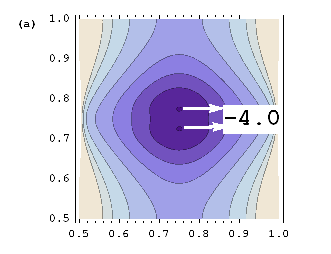}\includegraphics[totalheight=1.0 in]{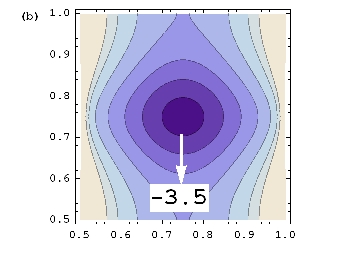}
\hspace{0in}%
\includegraphics[totalheight=1.0 in]{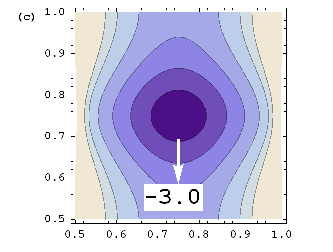}\includegraphics[totalheight=1.0 in]{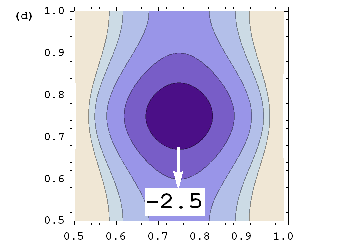}
\caption{(Color online) Contour plots the 2-D  $Re\{n\}$ as a function of (x,y)
dependent the probe detuning $\Delta_{p}$. All the parameters in (a) to (d)
are the same as in Figs. 2(a) to 2(d), respectively.\label{4}}
\end{figure}
Fig.3 shows the the three-dimensional plots for the $Re\{\mu_{r}\}$. It can be seen
from Fig.3 that $Re\{\mu_{r}\}$ has negative values under the same condition as Fig.2.
As shown in Fig.3 from (a) to (d), the increasing peak values of $Re\{\mu_{r}\}$
are observed when $\Delta_{p}$ was tuned to be $4.7\gamma$,
$5.0\gamma$, $5.3\gamma$ and $5.7\gamma$, in spite of the weaker magnetic response than
the electric response in Fig.2. What's more, we note that alone the x axis
the distribution pattern of $Re\{\mu_{r}\}$ shows the evolution of fin-like gradually
into spike-like. Then an increasing magnetic response is shown when $\Delta_{p}$ was
gradually tuned increasingly. The simultaneously negative permittivity and permeability
shown by Fig.2 and Fig.3 prove the 2-D left-handedness in the N-type four-level atomic system.

However, what interested us most is the value and distribution of refractive index in the x-y plane.
Fig.4 gives the contour plots for the real part of refractive index $Re\{n\}$ in the x-y plane.
The values indicated by the white arrows and distributions of the innermost contour lines corresponding to $Re\{n\}$ attract
our intention mostly. Two points are the innermost contour diagram in Fig.4(a), and they get the value being -4.0 for $Re\{n\}$ with their corresponding coordinates  $\{0.75\lambda, 0.72\lambda\}$ and
$\{0.75\lambda, 0.77\lambda\}$. The innermost contour line changes into one circle with its central
coordinate $\{0.75\lambda, 0.75\lambda\}$ in Fig.4(b), and the corresponding value of $Re\{n\}$
is -3.5. When $\Delta_{p}$ is tuned by $5.3\gamma$ and $5.7\gamma$,
the innermost contour lines are still homogeneous and isotropic circles with the same centre
coordinate $\{0.75\lambda,0.75\lambda\}$ in Fig.4(c) and Fig.4(d), respectively. And the values
of $Re\{n\}$ corresponding to the innermost contour lines are -3.0, -2.5 in Fig.4(c) and Fig.4(d),
respectively. The circle distributions with same centre coordinates in Fig.4(b),(c) and (d) manifest
that $Re\{n\}$ can have the same value in all directions. That's to say, $Re\{n\}$ is homogeneous and
isotropic in the x-y plane when $Re\{n\}$ gets the following values,-3.5, -3.0, -2.5. The coherent
N-type four-level atomic system can be a left-handed media with isotropic NRI when
it gets the above mentioned values in the x-y plane.

\section{Conclusion}
We suggested a new scenario for 2-D isotropic NRI via two orthogonal
standing-wave fields coupling two different
atomic transitions in the N-type atomic system.
In the x-y plane the simultaneously negative
permittivity and permeability exhibit, which
means the left-handedness in the N-type atomic system. What's more,
$Re\{n\}$=-3.5, -3.0, -2.5 are both homogeneous and isotropic in the
x-y plane in the same frequency band. Our scheme proposes a new scenario
of 2-D isotropic NRI which may give an impetus to some potential advantages or applications.
The proposed 2-D isotropic NRI media in our scheme should
be paid more attention from both theoretically and experimentally in the near future.

\end{document}